\documentclass[aps,prl,superscriptaddress,twocolumn,showpacs,preprintnumbers,amsmath,amssymb]{revtex4-1}
\usepackage{graphicx}
\usepackage{dcolumn}
\usepackage{bm}
\usepackage{upgreek}
\usepackage{txfonts}
\usepackage{color}
\newsavebox{\tempbox}

\begin{document}

\title{ Evidence for nonlocal electrodynamics in planar Josephson
junctions}


\author{A.~A. Boris}
\affiliation{Department of Physics, Stockholm University, AlbaNova
University Center, SE-106\,91 Stockholm, Sweden}
\author{A. Rydh}
\affiliation{Department of Physics, Stockholm University, AlbaNova
University Center, SE-106\,91 Stockholm, Sweden}
\author{T. Golod}
\affiliation{Department of Physics, Stockholm University, AlbaNova
University Center, SE-106\,91 Stockholm, Sweden}
\author{H. Motzkau}
\affiliation{Department of Physics, Stockholm University, AlbaNova
University Center, SE-106\,91 Stockholm, Sweden}
\author{A.~M. Klushin}
\affiliation{Institute of Physics of Microstructures, 603950
Nizhnij Novgorod, Russia}
\author{V.~M. Krasnov}
\affiliation{Department of
Physics, Stockholm University, AlbaNova University Center,
SE-106\,91 Stockholm, Sweden}

\date{\today }

\begin{abstract}
We study temperature dependence of the critical current modulation
$I_\mathrm{c}(H)$ for two types of planar Josephson junctions: a
low-$T_\mathrm{c}$ Nb/CuNi/Nb and a high-$T_\mathrm{c}$
YBa$_2$Cu$_3$O$_{7-\delta}$ bicrystal grain-boundary junction. At
low $T$ both junctions exhibit a conventional behavior, described
by the local sine-Gordon equation. However, at elevated $T$ the
behavior becomes
qualitatively different: the $I_\mathrm{c}(H)$ modulation 
field $\Delta H$ becomes almost $T$-independent and neither
$\Delta H$ nor the 
critical field for penetration of Josephson vortices vanish at
$T_\mathrm{c}$. Such an unusual behavior is in good agreement with
theoretical predictions for junctions with nonlocal
electrodynamics. We extract absolute values of the London
penetration depth $\lambda$ from our data and show that a
crossover from local to nonlocal electrodynamics occurs with
increasing $T$ when $\lambda(T)$ becomes larger than the electrode
thickness.

\end{abstract}

\maketitle

\begin{figure*}
     \centering
    \includegraphics[width=\textwidth]{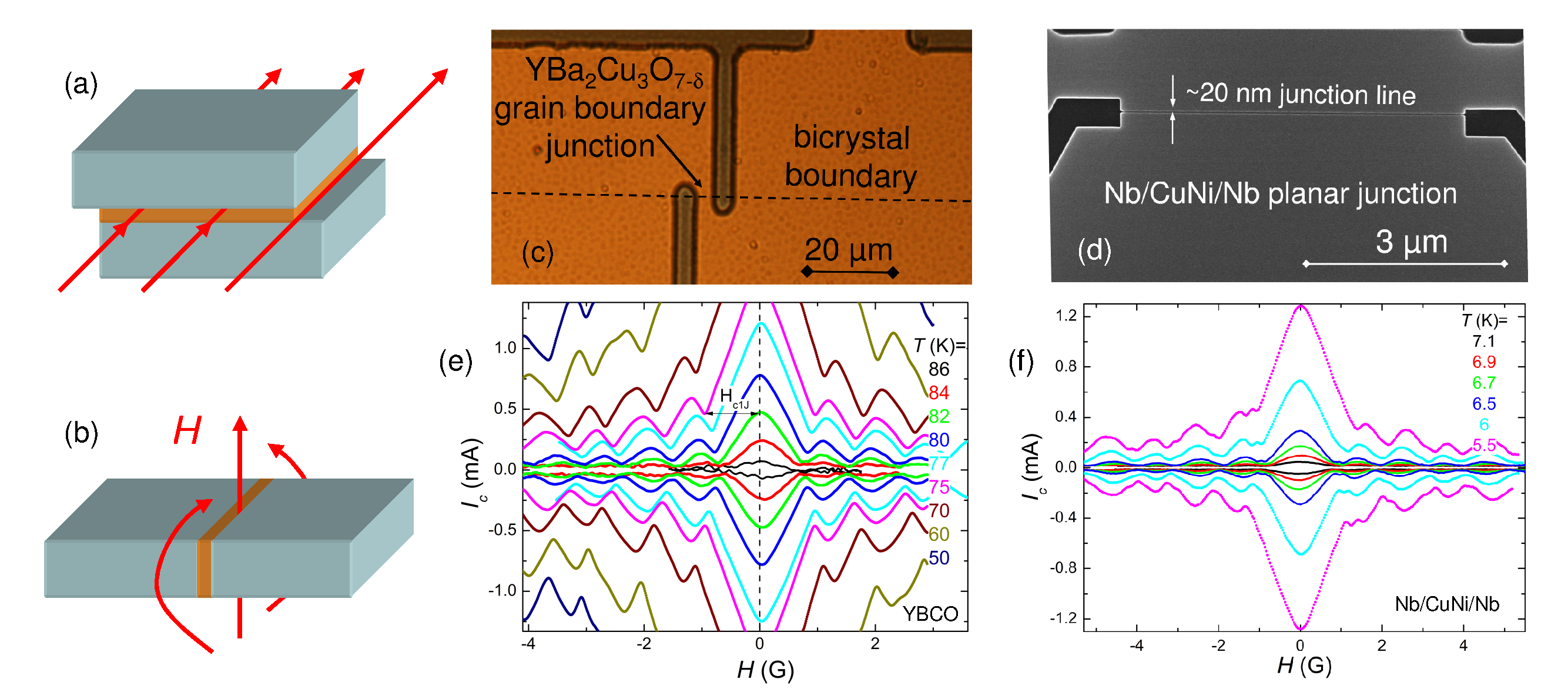}
\caption{(Color online). Geometries of (a) a conventional
overlap-type Josephson junction and (b) a planar Josephson
junction in applied magnetic field $H$. (c) Top view of the
studied YBCO bicrystal junction (optical microscope image). (d)
Top view scanning electron microscope image of the studied planar
Nb/CuNi/Nb junction. (e, f) $I_\mathrm{c}(H)$ modulation patterns
at different temperatures for (e) the YBCO junction and (f) the
Nb/CuNi/Nb junction.}
    \label{fig1}
\end{figure*}

Josephson junctions are usually formed by a barrier sandwiched
between two superconducting electrodes, as sketched in
Fig.~\ref{fig1}(a). Such overlap-type junctions have a local
electrodynamics, described by a differential sine-Gordon equation
\cite{Barone}. The locality is caused by the smallness of the
London penetration depth $\lambda$, at which the magnetic field is
screened in the electrodes, in comparison with the Josephson
penetration depth, $\lambda_\mathrm{J}$, at which the field is
varied along the junction, $\lambda_\mathrm{J} \gg \lambda$. In
this case the field is locked inside the junction. Its
distribution is quasi-one-dimensional and depends only on
local, instantaneous values of the Josephson phase difference. 
However, if the effective penetration depth becomes larger than
$\lambda_\mathrm{J}$, the magnetic field is no longer locked in
the junction. The field distribution becomes two-dimensional and
is determined by a nonlocal integro-differential equation
involving phases in the whole junction \cite{Ivanchenko}.

It has been suggested that nonlocal electrodynamics can be
realized in planar junctions formed at the edge between two
superconducting films with the thickness $d<\lambda$
\cite{Ivanchenko,Humphreys_1993,MintsJLTP1997,Kogan2001,Mints2008,Clem2011}.
Incomplete screening by thin films leads to increase of the
effective penetration depth. For $d\ll\lambda$ it is equal to the
Pearl length $\Lambda=2\lambda^2/d \gg \lambda$. Furthermore,
unlike in the case of overlap junctions, the field should be
applied perpendicular to the films. This leads to appearance of a
large demagnetization factor (flux focusing)
\cite{MintsJLTP1997,Rosenthal_1991,Humphreys_1993} and causes
spreading of stray magnetic fields at the surface of
superconducting electrodes to a distance of the order of the
junction width $w \gg \lambda$, as sketched in Fig.~\ref{fig1}(b).
In recent years several types of planar junctions have been
studied, including high-$T_\mathrm{c}$ grain boundary junctions
\cite{Winkler,Gross_Hc1_1993,Hilgenkamp_2002,Tafuri_2005} and
proximity-coupled junctions via semiconducting heterostructures
\cite{Takayanagi,Anticorrelation_2007}, ferromagnets
\cite{Anticorrelation_2007,CrO_Nature2006,Golod_PRL2010}, normal
metals \cite{Blamire,Ryazanov} or graphene \cite{HJLee}. The
effect of flux focusing has been established in previous works
\cite{Rosenthal_1991,Humphreys_1993,Gross_Hc1_1993}, but the role
of nonlocality remains to be clarified.

Theoretically it has been predicted that properties of nonlocal
and local junctions should be significantly different
\cite{MintsJLTP1997,Kogan2001,Mints2008,Clem2011}. The difference
is summarized in Table I. For example, in local junctions the
Josephson critical current $I_\mathrm{c}$ as a function of applied
magnetic field $H$ exhibits periodic-in-field Fraunhofer
modulation with a period $\Delta H\simeq \Phi_0/2 w\lambda$, where
$\Phi_0$ is the flux quantum. The $T$-dependence of $\Delta H$ is
determined by $\lambda(T)$. Close to $T_\mathrm{c}$, $\lambda(T)$
diverges and $\Delta H$ vanishes as $(T_\mathrm{c}-T)^{1/2}$. On
the other hand, for nonlocal junctions the $I_\mathrm{c}(H)$ is
not perfectly periodic in field and $\Delta H$ is determined by
spreading of stray fields, which depends solely on the geometry
$w$ and should be $T$-independent. Therefore, analysis of the
$T$-dependence of $I_\mathrm{c}(H)$ close to $T_c$ should provide
a clear distinction between the local and nonlocal models.

In this work we experimentally study the temperature dependence of
the $I_\mathrm{c}(H)$ modulation for two types of planar thin film
junctions: a low-$T_\mathrm{c}$ Nb/CuNi/Nb and a
high-$T_\mathrm{c}$ YBa$_2$Cu$_3$O$_{7-\delta}$ (YBCO) bicrystal
grain-boundary junction. We observe that at low $T$ junctions can
be described by local electrodynamics. However, at elevated $T$
both junctions exhibit a qualitatively different behavior,
consistent with occurrence of the nonlocal electrodynamics. We
show that a temperature driven crossover from local to nonlocal
electrodynamics, occurs when $\lambda(T)$ becomes larger than
electrode thickness.

\begin{table*}
\caption{\label{table1} Characteristic parameters of local
overlap-type and nonlocal planar Josephson junctions. Here
$d_\mathrm{eff}$ is the effective magnetic thickness of the
junction, i.e., the distance at which the magnetic field decays in
the electrodes, $\Delta H$ is the flux quantization field,
corresponding to the field interval of $I_\mathrm{c}(H)$
modulation, $\lambda_\mathrm{J}$ is the Josephson penetration
depth, $H_\mathrm{c1J}$ is the lower critical field for
penetration of Josephson vortices, $\Phi_0$ is the flux quantum,
$c$ is the speed of light, $d$ is the thickness of superconducting
films, $w$ is the junction width, $\lambda$ is the London
penetration depth and $I_\mathrm{c}$ is the critical current at
$H=0$.}
\begin{ruledtabular}
\begin{tabular}{cccccccc}
Junction type & Electrodynamics & $d_\mathrm{eff}$ & $\Delta H=\frac{\displaystyle \Phi_0}{\displaystyle w d_\mathrm{eff}}$ & $\lambda_\mathrm{J}$ & $H_\mathrm{c1J} \simeq \frac{\displaystyle 2\Phi_0}{\displaystyle \pi^2 \lambda_\mathrm{J} d_\mathrm{eff}}$ \\
overlap & Local & $\simeq 2\lambda $ & $\frac{\displaystyle \Phi_0}{\displaystyle 2 w \lambda}$ & $\displaystyle \sqrt{\frac{\displaystyle \Phi_0 c d w}{\displaystyle 16 \pi^2 \lambda I_\mathrm{c}}}$ & $\displaystyle \sqrt{\frac{\displaystyle 16\Phi_0 I_\mathrm{c}}{\displaystyle \pi^2 c d w \lambda} }$ \\
planar & Nonlocal & $\sim w$ & $\displaystyle \frac{1.8 \Phi_0}{\displaystyle w^2}$ & $\frac{\displaystyle \Phi_0 c d w}{\displaystyle 16 \pi^2 \lambda^2 I_\mathrm{c}}$ & $ \frac{\displaystyle 57.6 I_\mathrm{c} \lambda^2}{\displaystyle c d w^2}$ \\
\end{tabular}
\end{ruledtabular}
\end{table*}
\begin{figure*}[t]
   \centering
    \includegraphics[width=\textwidth]{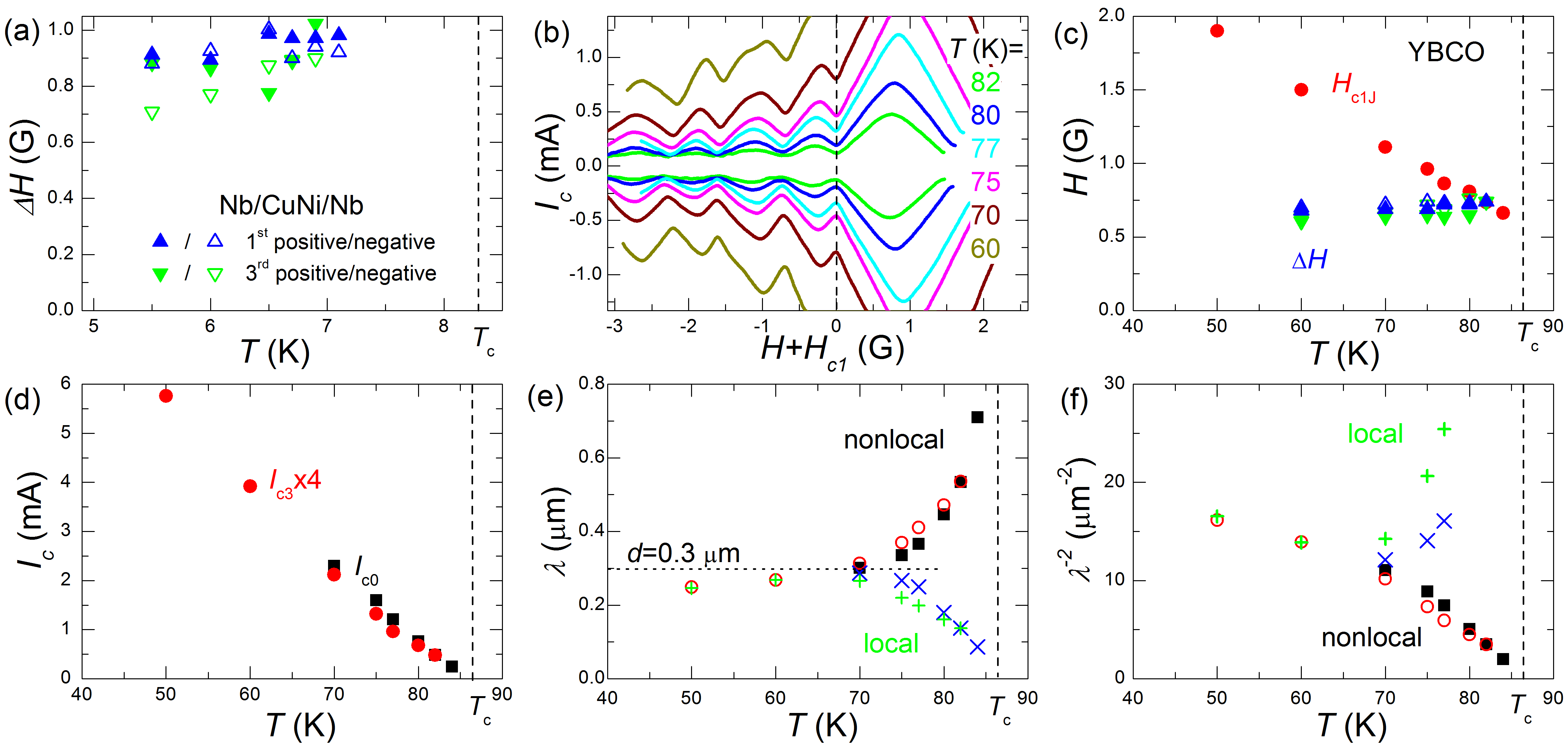}
\caption{(Color online). (a) Measured values of $\Delta H(T)$ for
the Nb/CuNi/Nb junction. Blue pointing up triangles represent the
width of the first $I_\mathrm{c}(H)$ lobe. Green pointing down
triangles represent the width of the third lobe. Data is shown for
both positive $H>0$ and negative $H<0$ fields for $I_\mathrm{c}
>0$. (b) Plots of $I_\mathrm{c}(H+H_\mathrm{c1J})$ at different
temperatures for the YBCO junction, with the first minimum at
$H<0$ shifted to the origin. Temperature independence of the flux
quantization field $\Delta H$ is clearly seen. (c) Measured values
of $\Delta H(T)$ (triangles) and $H_\mathrm{c1J}(T)$ (red circles)
for the YBCO junction. Colors of triangles represent the same
maxima of Fraunhofer patterns as in (a). (d) Measured dependence
$I_\mathrm{c0}(T)$ for the positive critical current at the
central maximum (black squares) and the maximum at the third lobe
$I_\mathrm{c3}(T)$ ($H<0$, positive current) scaled by a factor
four. (e, f) Temperature dependencies of $\lambda$ and
$\lambda^{-2}$ calculated using measured data, equations from
Table I, and $H_\mathrm{c1J}(T)$ and $I_\mathrm{c}(T)$ data from
panels (c) and (d). Squares and circles show results calculated
using equations for nonlocal planar junctions, crosses and pluses
- for local overlap-type junctions.} \label{fig2}
\end{figure*}
Figures~\ref{fig1} (c) and (d) represent images of studied
junction. The YBCO grain-boundary junction was fabricated on a
symmetrical bicrystal yttria-stabilized zirconia substrates ([001]
tilt) with a misorientation angle of 24$^{\circ}$. To prevent
interface reactions, $40\,\mathrm{nm}$ thick CeO buffer layer was
deposited prior to depositing the YBCO films. The YBCO film with
the thickness $d=300\,\mathrm{nm}$ and
$T_\mathrm{c}=86.5\,\mathrm{K}$ was grown epitaxially by reactive
high-oxygen-pressure metal co-evaporation using a rotating
substrate holder at Ceraco Ceramic Coating company \cite{ceraco}.
The substrate temperature was $665^{\circ}\mathrm{C}$ and the
deposition rate $\sim 0.4\,\mathrm{nm/s}$. Subsequently, a $w
\simeq 6\,\mathrm{\muup m}$ wide junction was patterned by
photolithography and cryogenic Ar$^+$-ion etching. Details of
bicrystal junction fabrication can be found in Refs.
\cite{Klushin1,Klushin2}. A low-$T_\mathrm{c}$ Nb/CuNi/Nb planar
junction ($T_\mathrm{c}=8.3\,\mathrm{K}$) was made by focused ion
beam (FIB) etching of a narrow ($\sim 20\,\mathrm{nm}$) grove
through a Nb/CuNi ($70/50\,\mathrm{nm}$) bilayer film. The films
were deposited at room temperature on oxidized Si substrates by
dc-magnetron sputtering at a base pressure $\sim 10^{-8}$ Torr and
processing Ar pressure 5 mTorr. The Cu$_{57}$Ni$_{43}$ film was
deposited by co-sputtering from Cu and Ni targets with controlled
Ni and Cu deposition rates. The bilayer film was patterned by
photolithography and ion etching (CF$_4$ reactive ion etching for
Nb and Ar-milling for CuNi). The width of the junction was $w
\simeq 5\,\mathrm{\muup m}$. Details of the Nb/CuNi/Nb planar
junction fabrication can be found in Refs.
\cite{Anticorrelation_2007,Golod_PRL2010}. Measurements were
performed in a cryogen-free cryostat using a four-probe
configuration. Magnetic field was applied perpendicular to the
films, as illustrated in Fig.~\ref{fig1}(b).

Figure~\ref{fig1}(e) and (f) represent measured $I_\mathrm{c}(H)$
modulation patterns at different temperatures for the YBCO and the
Nb planar junctions, respectively. At low $T$, the central maxima
at $H=0$ are significantly wider than subsequent lobes in
$I_\mathrm{c}(H)$ and exhibit a characteristic linear decrease
with field. Such behavior is typical for long junctions,
$w>4\lambda_\mathrm{J}$ \cite{Barone,Gross_Hc1_1993,NonUn}. In
this case the external magnetic field can be screened within the
junction. The linear-in-field central maximum corresponds to the
Meissner state without Josephson vortices inside the long
junction. It ends at the Josephson lower critical field
$H_\mathrm{c1J}$ \cite{Gross_Hc1_1993}. At $H>H_\mathrm{c1J}$
Josephson vortices penetrate the junction and the
$I_\mathrm{c}(H)$ modulation is restored. Thus defined
$H_\mathrm{c1J}$ is marked by the horizontal arrow for the curve
at $T=75\,\mathrm{K}$ in Fig.~\ref{fig1} (e). For the Nb junction,
Fig.~\ref{fig1}(f), the linear-in-field central maximum is seen
only at the lowest temperatures. At elevated $T$ the
$I_\mathrm{c}(H)$ starts to resemble a Fraunhofer pattern,
characteristic for short junctions, $w \lesssim
\lambda_\mathrm{J}$. However, that $I_\mathrm{c}(H)$ modulation
for short nonlocal junctions deviates from the Fraunhofer
modulation with constant flux quantization field $\Delta H$,
typical for overlap junctions \cite{Barone}. Indeed, from
Fig.~\ref{fig1}(f) it can be seen that at high $T$ the first
minima (half the width of the central maximum) is narrower than
the subsequent lobes: $H_1/\Delta H \simeq 0.83$. This is close to
the theoretical value $0.8173$, calculated by J.R.~Clem for short
planar junctions \cite{Clem2011}. The same ratio $H_1/\Delta H
\simeq 0.83$ is observed for the YBCO junction very close to
$T_\mathrm{c}$, when the junction becomes short, see the curve at
$T=84\,\mathrm{K}$ in Fig.~\ref{fig1}(e).

Triangles in Figs.~\ref{fig2}(a) and (c) represent the
$T$-dependence of $\Delta H$ for Nb and YBCO junctions,
respectively, obtained from the widths of the first and the third
side-lobes of $I_\mathrm{c}(H)$ at positive (filled) and negative
(open symbols) fields. It is seen that $\Delta H (T)$ is almost
constant. It does not show a tendency to vanish at $T\rightarrow
T_\mathrm{c}$, as expected for local junctions, but rather even
slightly increases with increasing $T$. The absolute value $\Delta
H \sim 1~\mathrm{Oe}$ is consistent with the prediction from
Table~I for nonlocal planar junctions with $w=5-6\,\mathrm{\muup
m}$. To clearly see the $T$-independence of $\Delta H$, we replot
the $I_\mathrm{c}(H+H_\mathrm{c1J})$ data for the YBCO junction in
Fig.~\ref{fig2}(b), offset by $H_\mathrm{c1J}(T)$ so that the
first minimum at $H<0$ is shifted to the origin (dotted vertical
line). From this plot it is clearly seen that the field scale of
$I_\mathrm{c}(H)$ modulation is indeed almost $T$-independent.

Circles in fig.~\ref{fig2}(c) represent the temperature dependence
of the critical field $H_\mathrm{c1J}(T)$ for the YBCO junction.
It linearly decreases with increasing $T$, but, remarkably, does
not vanish at $T\rightarrow T_\mathrm{c}$, as would be expected
for conventional local Josephson junctions. As follows from Table
I, the $H_\mathrm{c1J}(T)$ depends on $I_\mathrm{c}(T)$ and
$\lambda(T)$, but the functional dependence is different for the
local and the nonlocal cases. Fig.~\ref{fig2}(d) shows the
$T$-dependence of the critical current for the YBCO junction. It
was obtained for the central maximum at zero field $I_{c0}$ and
the third side-lobe maximum $I_{c3}$ (scaled by a factor 4).

Using the measured $H_\mathrm{c1J}(T)$, $I_\mathrm{c}(T)$ and the
expressions for $H_\mathrm{c1J}$ from Table~I we calculate the
$\lambda(T)$ dependence. 
Figs.~\ref{fig2} (e) and (f) represent the obtained $\lambda(T)$
and $\lambda^{-2}(T)$ for overlap (local) \cite{Note} and planar
(nonlocal) cases. It is seen that at low $T$ there is no
qualitative difference between the two models, which is probably
the reason why a clear distinction could not be made from previous
similar studies \cite{Humphreys_1993,Gross_Hc1_1993}. However,
such a distinction becomes apparent from our data at elevated
temperatures. From Figs.~\ref{fig2} (e) and (f) it is seen that at
$T>70\,\mathrm{K}$ the local theory provides totaly erroneous
results with vanishing $\lambda(T)$ at $T\rightarrow
T_\mathrm{c}$. To the contrary, the nonlocal theory provides both
a quantitatively correct absolute value of $\lambda(0) \sim
0.2\,\mathrm{\muup m}$ at low $T$ and a qualitatively correct
$T$-dependence at $T\rightarrow T_\mathrm{c}$ with diverging
$\lambda(T\rightarrow T_\mathrm{c})$ and linearly vanishing
$\lambda^{-2}(T) \propto T_\mathrm{c}-T$
\cite{Lambda_muSR,Prozorov}. This provides a clear evidence for
nonlocal electrodynamics. The dotted horizontal line in
Fig.~\ref{fig2} (e) demonstrates that the divergence between local
and nonlocal models occurs when $\lambda(T)$ becomes larger than
the electrode thickness $d=300$ nm. This indicates that a
temperature-driven crossover from local to nonlocal
electrodynamics takes place.

Now we can understand why $H_\mathrm{c1J}$ does not vanish at
$T\rightarrow T_\mathrm{c}$ in our planar junctions. As seen from
Table~I, in the nonlocal model $H_\mathrm{c1J}(T) \propto
I_\mathrm{c}(T)\lambda^2(T)$. From Figs.~\ref{fig2} (d) and (f) it
is seen that $I_\mathrm{c}(T) \propto \lambda^{-2}(T) \propto
T_\mathrm{c}-T$ close to $T_\mathrm{c}$. Therefore,
$H_\mathrm{c1J}$ is determined by the ratio of two similar
vanishing functions and remains finite at $T_\mathrm{c}$. This is
an intrinsic property and a consequence of nonlocal the
electrodynamics in planar thin-film Josephson junctions.

To conclude, we have studied the temperature dependence of the
critical current modulation for low-$T_\mathrm{c}$ and
high-$T_\mathrm{c}$ thin film planar Josephson junctions. We
observed a temperature-driven crossover from local to nonlocal
electrodynamics. It takes place when $\lambda(T)$ becomes larger
than the electrode thickness $d$. At elevated temperatures both
junctions exhibited a similar unusual behavior, which is in
drastic discrepancy with that for conventional overlap-type
junctions, described by the local sine-Gordon equation: (i) The
flux-quantization field $\Delta H$ of $I_\mathrm{c}(H)$ modulation
was $T$-independent, did not vanish at $T_\mathrm{c}$ and was
determined not by the London penetration depth $\lambda$ but
solely by the junction geometry $w$. (ii) The critical field for
penetration of Josephson vortices $H_\mathrm{c1J}$ remained finite
at the critical temperature $T_\mathrm{c}$. These observations
provided clear evidence for nonlocal electrodynamics in thin film
planar Josephson junctions in good agreement with theoretical
predictions \cite{MintsJLTP1997,Kogan2001,Mints2008,Clem2011}. Our
results indicate that the nonlocality indeed deeply affects
properties of planar junctions.

The work was supported by the Ministry of Education and Science of
Russian Federation project Nr.~8837 and the STINT foundation. We
are grateful to R.G. Mintz for valuable remarks and to the Core
facility in Nanotechnology at Stockholm University for technical
support.


\begin{references}
\bibitem{Note} In the field perpendicular to films
the expression for $H_\mathrm{c1J}$ for overlap junctions should be
corrected by the corresponding demagnetization factor,
$H_\mathrm{eff}=H/(1-D)$. We have chosen $D=0.91$ to obtain the correct
absolute value of $\lambda$ at low-$T$. This value of $D$ is
reasonable for our thin-film junctions.
\end{references}
\end{document}